# Electrical resistivity of microstructural components in Al-Mg-Si alloys

Gautam Kumar [1], Amram Azulay [1], Omer Coriat [1], and Hanna Bishara [1,*]

[1] Department of Materials Science and Engineering, The Iby and Aladar Fleischman Faculty of Engineering, Tel Aviv University, Tel Aviv, 6997801, Israel

[*] Corresponding author: hbishara@tauex.tau.ac.il

**Abstract**

Al-Mg-Si alloys are utilized in large scale electrical conduction applications thanks to their low density, high strength, and low electrical resistivity. The alloying elements, Mg and Si, are introduced to improve the mechanical strength; however, the formed defects also suppress electrical conductivity, adversely affecting the material performance. Here, we investigate the impact of alloying, heat treatment, and the corresponding microstructure, on the electrical resistivity of overaged alloys having 0.5-9.5 at. % solute. The crystal structure, composition, and microstructure are characterized by X-ray diffraction, energy dispersive X-ray spectroscopy, electron backscatter diffraction, and scanning electron microscopy. The electrical resistivity of the microstructural components, i.e., the Al solid solution matrix and the Si and $Mg_2Si$ precipitates, are directly measured using a microscale four-point probe setup inside a scanning electron microscope. We find that the Al solid solution matrix is up to 15 % more resistive than pure Al, depending on the heat treatment rather than the composition, and that regions including Si or $Mg_2Si$ precipitates are equally resistive. Additionally, the bulk alloy resistivity, measured conventionally on a macroscopic length scale, increases linearly up to 60 % with increasing total solute concentration up to ~ 10 at. %. This study relates the electrical resistivity of Al-Mg-Si alloys, measured at microscopic and macroscopic length scales, with their microstructure and composition.

**Highlights**

- SEM in-situ electrical experiments are coupled with microstructure characterization
- Grain interior and individual precipitates resistivity were directly measured
- Grain interior resistivity is affected by heat treatment, not composition
- Si and $Mg_2Si$ precipitates contribute equally to the resistivity
- Alloy resistivity is quantified by precipitates volume and annealing condition

**Graphical Abstract**

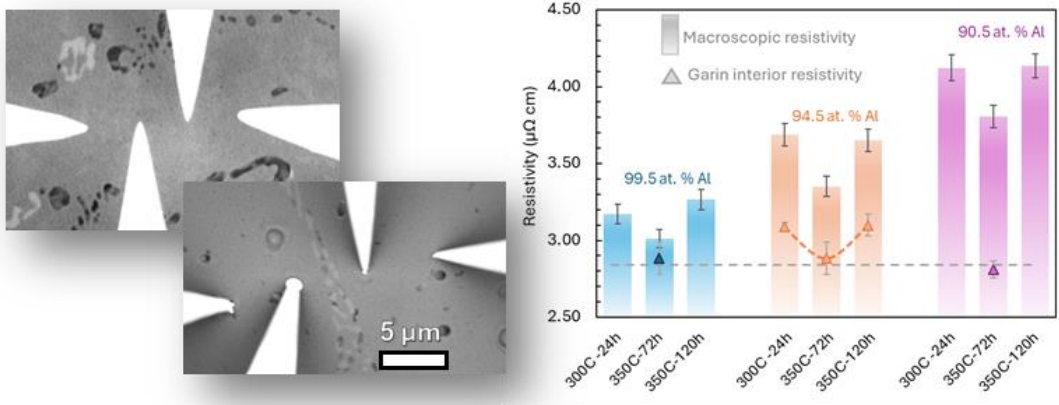

# 1 Introduction

Aluminum alloys are promising substituents for copper-based materials in large scale electric transmission applications thanks to their low weight and low electrical resistivity [1–5]. Even though the electrical conductivity of Al is about 63 % than that of Cu, its mass density is about one-third [6], rendering the resistivity per mass ratio attractive. Among the methods available to achieve the mechanical strength required for applications such as hanging transmission lines is alloying with Mg and Si, which enable precipitation hardening [2,3]. Though crystal defects usually improve material strength, they also impede electron transport, thus increasing electrical resistivity. Therefore, the challenge at hand is to obtain a sufficiently high strength with a minimal increase of electrical resistivity [8–11].

Usually, electrical characteristics of solids are measured at macroscopic length scales (mm to cm). Thus, contributions of individual microstructural components, such as precipitates and grain boundaries, are included in the interaction volume and hence their contributions to the measured value are not resolved directly. To quantify their specific resistivities, Matthiessen rule and physical models [9,12–15] are employed. For instance, the resistivities of grain boundaries and dislocations in Al-Mg-Si alloys were evaluated as 2.6 p$\Omega$ cm$^2$ and 2.71×10$^{-25}$ $\Omega$ m$^3$, respectively [2]. Furthermore, it was shown that the specific resistivity of the Al solid solution increases with increasing concentration of alloying elements, independently of heat treatments [2,8,16]. The same defect accumulative approach is adopted to unravel the defect-resolved mechanical properties in Al-Mg-Si, i.e., determine the total strength by strength contributed by solid solution, precipitates, grain boundaries, and dislocations. [2,17,18].

In Al-Mg-Si alloys, electrical resistivity is optimized by designing the microstructure through heat and mechanical treatments. For example, the resistivity of Al 6201 alloy was reported to decrease with the increasing the heat treatments duration due to evolving coherency, amount, and shape of nanoscale precipitates [17,19]. Specifically, alloys with a Si and Mg concentration of ~ 1 at. % attain a maximal resistivity of ~ 3 μ$\Omega$ cm after heat treatments of tens of hours at 180 °C [17]. The Mg/Si ratio impacts the alloy resistivity through phase evolution kinetics and aging, affecting the precipitation state [18,20]; for example, Si excess provides a lower resistivity compared to Mg excess due to segregation effects [18]. The electrical resistivity of Al-Mg-Si alloys linearly increases with increasing solute concentration up to 0.8 at.

% [20] and with precipitate density [2,19]. However, resistivity measurements of the microstructural defects in Al-Mg-Si alloys are still missing. Therefore, decoupling the Al matrix from the precipitate's resistivity is called for. This decoupling approach was successfully applied in Cu-based alloys [21] and full and half Heusler alloys [22,23], to elucidate conduction mechanisms.

In this work, we employ a microscale four-point probe setup inside a scanning electron microscope (SEM) to explore the relation between electrical resistivity and microstructure in Al-Mg-Si alloys. Here, we consider overaged alloy to enable stabilization of the non-coherent phase secondary; though decreasing strength but allowing higher ductility and thermal stability for elevated temperature service. This novel technique allows us to distinguish between the resistivity of Al solid solution grains and regions including Si and $Mg_2Si$ precipitates, in alloys having different compositions and that underwent different heat treatments.

## 2  Experimental

### 2.1  Synthesis, processing, and structural characterization

Three Al-Mg-Si ingots with compositions described in **Table 1** were prepared by melting appropriate amounts of master alloys having nominal compositions of $Al_{99}Mg_{0.5}Si_{0.5}$ and $Al_{90}Mg_5Si_5$ (HMW Hauner GmbH & Co. KG, Germany) in an induction melting furnace (MCV100, Indutherm Erwärmungsanlagen GmbH, Germany). The paper mainly discusses the samples presented in **Table 1**; however, additional ones having different compositions are considered in the last section. The melting process included initial venting of the chamber to 0.1 mbar, purging three times with Ar gas (99.999 % plus $H_2O/O_2$ purification), and heating to 730°C at 500 mbar. An additional purge was performed upon melting to clear trapped gases. The liquid was poured into a cylindrical graphite mold, pre-heated to 350°C, and left to cool inside the furnace. The as-cast ingots were diced into samples using a wire saw (DWS.100, Diamond WireTec GmbH & Co.KG, Germany). The cut samples were heat treated in a vacuum furnace at three different conditions: 24 h at 300°C, 72 h at 350°C, and 120 h at 350°C. Such annealing condition transforms all precipitates into the stable non-coherent state, though lowering strength, but stabilizing microstructure for elevated temperature application. Finally, the samples were polished with a final step of and colloidal silica 50 nm suspension, and then vibro-polished for 2 h with the same suspension (*Qpol* 250 M1 and *Qpol Vibro*, ATM Qness GmbH, Germany).

**Table 1**. Compositions (at. %) of the Al-Mg-Si samples evaluated by energy-dispersive X-ray spectroscopy (EDS). Uncertainties are considered as 1 at. %.

| Sample # | Al | Si | Mg |
|---|---|---|---|
| 1 | 99.5 | < 1 | < 1 |
| 2 | 94.5 | 2.5 | 3 |
| 3 | 90.5 | 5.5 | 4 |

Crystal structure and phase identification were carried out using X-ray diffraction (XRD), employing a D8 diffractometer (Bruker, USA) for an angular range of 20-100° (2θ) with 0.025° sampling resolution. Microstructure and elemental compositions were analyzed by a *Phenom XL G2* SEM (Thermo Fisher Scientific, USA) equipped with an energy-dispersive X-ray spectroscopy (EDS) detector; backscattered electrons were recorded for imaging to resolve phase contrast. The relative abundance of grain boundaries compared to phase boundaries was revealed by electron backscatter diffraction (EBSD) analysis using a *Quanta* 200 FEG SEM (Thermo Fisher Scientific, USA) equipped with a *NORDLYS* II detector (Oxford Instruments, UK).

## 2.2 Electrical characterization

Electrical resistivity was measured on both macroscopic and microscopic length scales. The macroscopic (bulk, global) electrical resistivity was measured by the Van der Pauw method at room temperature (HFS 600, INSTEC Inc., USA) applying a current of 50 mA in a sequence of hundred DC pulses; the voltage was measured at the center of each pulse, and the resistivity was calculated from the average of the hundred pulses. This process was repeated at least twice for each sample to confirm consistency, and the reported resistivity is the mean value of these equivalent measurements.

On the microscale, electrical resistivity was measured at room temperature by the four-point probe method using a probe station (PS4, Kleindiek Nanotechnik GmbH, Germany) inside the *Phenom XL G2* SEM. Electrical modulations protocols for the measurements are the same as in Van der Pauw measurements described above. To resolve the grain interior resistivity, consisting of Al saturated solid solution and nanoprecipitates, a linear configuration of the probes was conducted, in which all four terminals are located within the same grain and no precipitates are observed in between. The spacing between the terminals was set to 5 μm. Four different locations were inspected to ensure repeatability and minimal impact of any macro-precipitates beneath the surface in the measured region. Yet, the contributions of precipitates to resistivity were determined by applying a similar four-point probe setup, where individual precipitates were positioned in between the voltage terminals. Precipitates of the two secondary phases, Si and $Mg_2Si$, having different morphologies, were inspected. Measurement protocols, accounting for non-uniform distance between probes, methodology, and error propagation are described elsewhere [24].

## 3 Results and Discussion

### 3.1 Microstructure

**Figure 1** shows XRD patterns collected from samples heat treated at 350°C for 72 h, indicating reflections of Al solid solution and Si, and $Mg_2Si$, corresponding to ICDD PDF #04-012-7848, 04-007-5232, and 01-091-0672, respectively. The peak at ~ 37° is an artifact due to Ni filter used in the measurement. The rest of the samples, treated at different temperatures and durations, show similar patterns with varying peak intensities (not presented). The positions of Al are shifted towards lower angles compared to the reference, indicating a larger lattice compared to the elemental metal. The lattice parameters obtained range from 0.45164 to 0.405247 nm (acquired by Pawly method), depending on composition and heat treatments. No clear correlation was further found between the lattice parameter and electrical resistivity of grain interior (to be further discussed).

The relative peak intensities obtained for all the samples deviate from that of the powder diffraction, indicating either a solidification texture, extremely large grains, or both. Indeed, EBSD measurements (not presented) confirm hundreds of micron-sized grains. These large grains are observed as contrast variations in the matrix, imaged using backscattered electrons signal, see **Figure 2**. Nevertheless, the crystallographic texture is irrelevant to the current work since the matrix exhibits a face centered cubic lattice having an isotropic electrical conductivity tensor. The influence of grain size and distribution of grain boundaries type on the electrical properties of the inspected alloys are negligible due to low density of grain boundaries, as discussed in section 3.2.

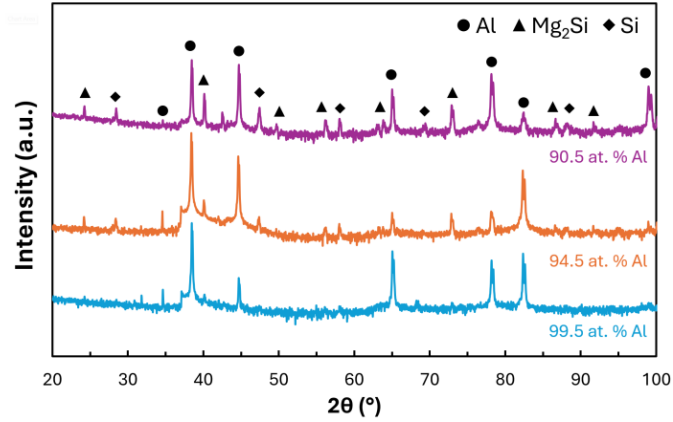

**Figure 1**: X-ray diffraction patterns collected from Al-Mg-Si samples with 90.5, 94.5, and 99.5 at. % Al, heat treated at 350°C for 72 h, matching the Al, Mg$_2$Si, and Si ICDD PDFs mentioned in the text. Note that the peak at ~ 37° is attributed to the nickel filter.

The microstrcuture of alloys with varying compositions and heat treatements is presented in **Figure 2**. Micron and sub-micron sized Si and Mg$_2$Si phases are present within the Al matrix, appearing as bright and dark features in the SEM micrographs, respectivley. **Figure 2(a)-(c)** show micrographs of alloys having the same composition, 94.5 at. % Al, heat treated at different conditions . The precipitates are initially formed in a eutectic microstructure, as predicted by phase diagrams, see [25–27]. Longer heat treatment durations leads to a partial spherodization of the precipitates lamellar structure. Moreover, it is shown in **Figure 2(d),(b),(e)** that the volume fraction of the secondary phases increases with increasing Si and Mg content, as expected. The slight contrast variations in the Al matrix originate from electron channeling, indicating large grains with different orientations, confirming the abovementioned XRD and EBSD findings.

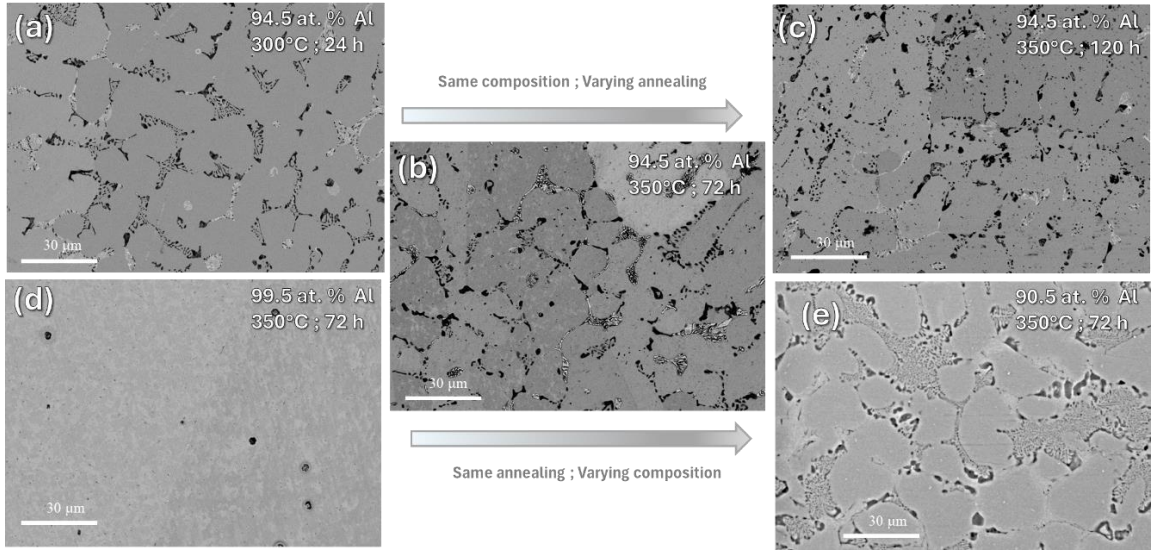

**Figure 2**: Microstructure of the Al-Mg-Si alloys. (a)-(c) Scanning electron microscopy backscattered electrons micrographs of samples with 94.5 at. % Al, heat treated at the conditions specified. (d), (b), and (e) show micrographs of samples with 99.5, 94.5, and 90.5 at. % Al, heat treated at the same condition.

**Figure 3** presents an example of elemental mapping obtained from EDS data of the 94.5 at. % Al alloy, confirming that Si- and Mg-rich regions do not always overlap. Corroborating the XRD analysis, this suggests two types of precipiates: Si and $Mg_2Si$. The elemental mapping is a key technique to identify the precipitates to be invetigated by microscale electrical measurements.

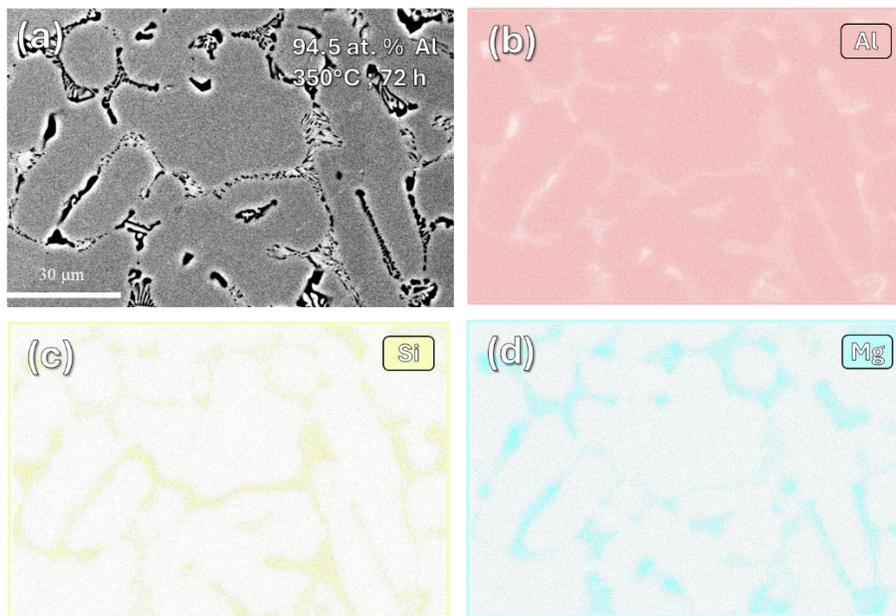

**Figure 3**: Elemental maps generated for the 94.5 at. % Al alloy by energy-dispersive X-ray spectroscopy using characteristic x-ray photons of (b) Al, (c) Si, and (d) Mg collected from the region shown in (a).

## 3.2 Bulk resistivity

**Figure 4** summarizes the bulk resistivities of the alloys. It is shown that reduction in Al content, *i.e.*, increasing Si and Mg content, leads to a general trend of an increase in the bulk resistivity. Interestingly, we find that the heat treatments applied here affect the samples of each composition similarly; that is, a heat treatment at 350°C for 72 h yields the lowest resistivity for all compositions.

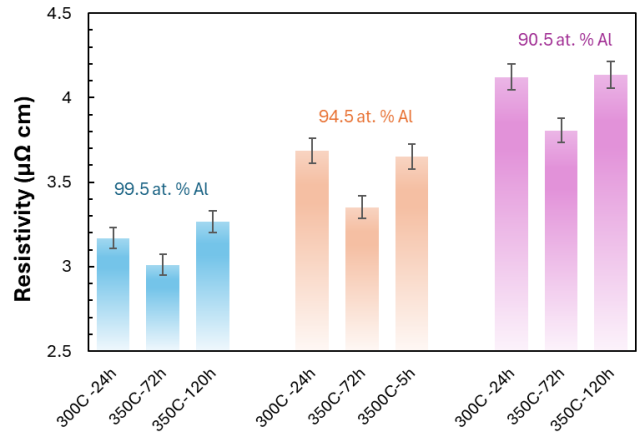

**Figure 4**: Bulk electrical resistivity values measured for the Al-Mg-Si samples; compositions and heat treatments are noted.

To explain the bulk resistivity trends observed in **Figure 4**, the contributions of microstructural components must be resolved. The superposition of microstructural components resistivity yields the bulk electrical resistivity, $\rho$, as described by Matthiessen rule, see equation (1).

$$\rho = \rho_0 + C_{sol}\rho_{sol} + V_{pr}\rho_{pr} + C_V\rho_{vac} + L_{dis}\rho_{dis} + S_{GB}\rho_{GB} \tag{1}$$

Here, $\rho_0$ is the matrix resistivity, $C_{sol}$ is the molar fraction of the alloying elements in the matrix, $V_{pr}$ is the precipitates volume fraction, $C_V$, $L_{dis}$, and $S_{GB}$ are the densities of vacancies, dislocations, and grain boundaries, respectively, and $\rho_{sol}$, $\rho_{pr}$, $\rho_{vac}$, $\rho_{dis}$, and $\rho_{GB}$ are the specific resistivities of the alloying elements in the solid solution, precipitates, vacancies, dislocations, and grain bounders, respectively.

In this work, we neglect the contribution of vacancies since samples were treated at a small temperature range and did not undergo mechanical treatments. Also, in the absence of plastic deformation, we assume low dislocation density and thus a negligible contribution to resistivity. Additionally, thanks to an average grain size of roughly a hundred micrometers, assessed by EBSD, we neglect the contribution of grain boundaries to resistivity. In fact, the contributions of vacancies, dislocations, and grain boundaries become significant in Al-based alloys above a concentration of 0.5 at. %, density of $10^{15}$ m$^{-2}$, and grain size of ~ 10 μm, respectively [2]. Therefore, we conclude that the main components contributing to the electrical resistivity of the alloy are Al matrix grains (solid solution) and precipitates of secondary phases.

The solubility limits of Si and Mg in the ternary Al-Si-Mg system, at the temperatures applied in this study, are in the order of tenths of an at. % [25], which is well below the added amounts. Therefore, the high solute content yields an increased precipitates density, as shown in **Figure 2(d)**, **(b)**, and **(e)**, and hence a higher resistivity. It should be noted that we consider only precipitates of micron and sub-micron length scales. The nanoscale ones, usually observed inside Al grains in Al-Mg-Si alloys [8,17,28], are treated as a part of the Al matrix, thus their resistivity is included within the grain interior.

### 3.3    Grain interior resistivity

**Figure 5(a)** shows an example of microscale measurements carried out within a grain interior; the resistivity values are reported in **Figure 5(b)** as black marks. The bulk resistivities are presented in the same figure as bars for comparison. We find that alloys with 99.5, 94.5, and 90.5 at. % Al that underwent the same heat treatment at 350°C for 72 h exhibit similar grain interior resistivity, namely, 2.88 ± 0.10, 2.88 ± 0.11, and 2.81 (± 0.05) μΩ cm, respectively. This result implies that the concentration of Mg and Si in the grains is similar, and the nano-precipitates evolution achieved similar morphologies. In contrast, for a fixed composition, the grain interior resistivity depends on the heat treatment, as seen for the 94.5 at. % Al sample. This result is attributed to heat-treatment driven variations in the nano-precipitate state and composition of the solid solution, and hence, the associated resistivities [8,11]. No clear correlation was found between the matrix lattice parameters (extracted from XRD analysis) and grain interior resistivity, suggesting that not only lattice distortions due to point defects affect grain interior resistivity, but also nano-precipitates play a significant role.

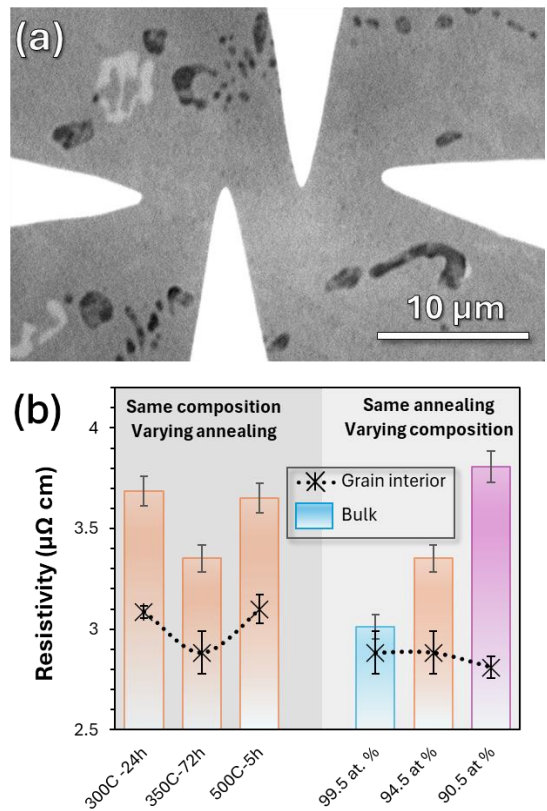

**Figure 5**: (a) Scanning electron microscopy backscattered electrons micrograph of a microscale electrical measurement on an Al grain interior region. (b) Grain interior resistivity measured for alloys. Error bars represent the standard deviation of 3-4 measurements. The left side relates to alloys having the same composition, and that underwent94.5 at. % Al the noted heat treatment. The right side refers to alloys that underwent the same heat treatment,350°C for 72 h, and have the indicated composition. The bar chart shows the bulk resistivities of the alloys, for comparison.

### 3.4 Precipitate resistivity

To inspect the resistivity across Si and $Mg_2Si$ precipitates having different shapes, microscale electrical measurements were employed. We resolved the resistivity of the Si and $Mg_2Si$ precipitates in alloys having 94.5 and 90.5 at. % Al, heat treated at 350°C for 72 h. **Figure 6(a)-(e)** presents a few of the measurement probes configurations for measuring the different microstructural components, i.e., Si and $Mg_2Si$ precipitates, having a bright and dark contrast, respectively, in different morphologies. Since the images provide a two-dimensional projection, each measurement was carried out in three locations having a similar microstructural component, thus validating the results. The value reported per component is the average of these three measurements. We note that as the measurement is performed across precipitates, the interaction volume of the applied electric field includes the inspected precipitate and its adjacent Al solid solution matrix. To allow for a proper comparison between the precipitates, similar inter-terminal spacing and precipitate thickness were chosen for the measurements.

**Figure 6(f)** summarizes the average resistivity values measured across all individual precipitates. We find that within the interaction volume of the measurement, precipitates having an average projected thickness of 700 nm increase the resistivity by ~ 55 % with respect to pure Al [2]. Interestingly, the same resistivity, within measurement uncertainties, is obtained regardless of the precipitate type, Si or $Mg_2Si$, and shape, indicating a negligible effect of the precipitate identity on the effective resistivity. As a matter of fact, the measured resistivity corresponds to series connection of grain interiors and a precipitate. The resistivity of the Al solid solution matrix ~ 2.5 μΩ cm is three orders of magnitude lower than that of both secondary phases ~ $10^3$ μΩ cm[29,30], respectively. Therefore, a negligible effect of the precipitate identity is expected, resulting in a similar effective resistivity for regions including either phase. This observation implies that the volume fraction of the secondary phases, rather than their identity, is the major cause for the increase in resistivity with increasing solute content or equivalently precipitates volume. In the following paragraph, we quantify this effect.

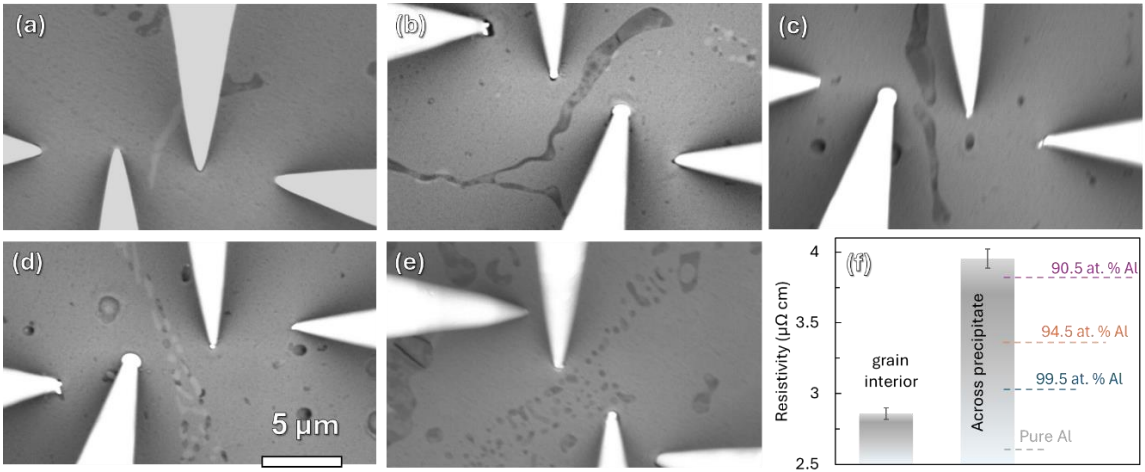

**Figure 6**: (a-e) Scanning electron microscopy backscattered electrons micrographs of microscale four-point probe configurations across Si and $Mg_2Si$ precipitates, appearing as bright and dark features, respectively.

A variety of precipitate shapes and microstructures were inspected. (f) Average resistivity values measured across eight precipitates. Error bar represents the standard deviation of the measured values. Experiments relate to Al-Mg-Si samples with 94.5 and 90.5 at. % Al, heat treated at 350°C for 72 h. Grain interior and bulk resistivities (right side, dashed lines) are shown for comparison.

The bulk resistivity is described by the sum of the resistivities of the microstructural components, see equation (1). Then, for a specific heat treatment, the resistivity of the Al-Mg-Si alloy can be described as a combination of the resistivity of pure aluminum, the saturated solid solution (directly measured as grain interior resistivity), and the precipitates. To quantify the precipitates' contribution, we inspect the bulk resistivity with respect to the solute concentration, both Si and Mg, as shown in **Figure 7**. The excess solute forms microscale precipitates [26], providing additional electron scattering centers, and their volume fraction increases with increasing solute content. Therefore, bulk resistivity linearly depends on the solute content, $x_{solute}$, as described by equation (2), which is Matthiessen rule (equation 1) in our case. We note that the grain interior consists of a solid solution and nano-scale precipitates, which are treated as one term.

$$\rho = \rho_0 + \rho_{sol}^{sat} + a \cdot x_{solute} = b + a \cdot x_{solute} \ [\Omega] \qquad (2)$$

Here, $a$ is constant proportional to the precipitate resistivity, and $b$ denotes the sum of pure Al and saturated solid solution resistivities, $\rho_0$ and $\rho_{sol}^{sat}$, respectively. Accordingly, the linear relation between resistivity and solute content is valid for alloys having the same grain interior resistivity (that is, the same value of $b$), which is a function of heat treatment.

**Figure 7(a)** shows a linear increase in resistivity with solute concentration for alloys treated at 350°C for 72 h. It is shown that the intersection with the vertical axis is the grain interior resistivity reported in **Figure 4**. Similar linear fits are obtained for alloys treated at 300°C for 24 and 350°C for 120 h, as shown in **7(b)**. In the latter cases, the grain interior resistivity is equal (within the measurement uncertainty), but different from the former case, as demonstrated in **Figure 4**. The linear regression coefficients are summarized in **Table 2**. Note that **Figure 7** includes two additional alloys with different compositions than those presented above; these were prepared to corroborate the linear relation.

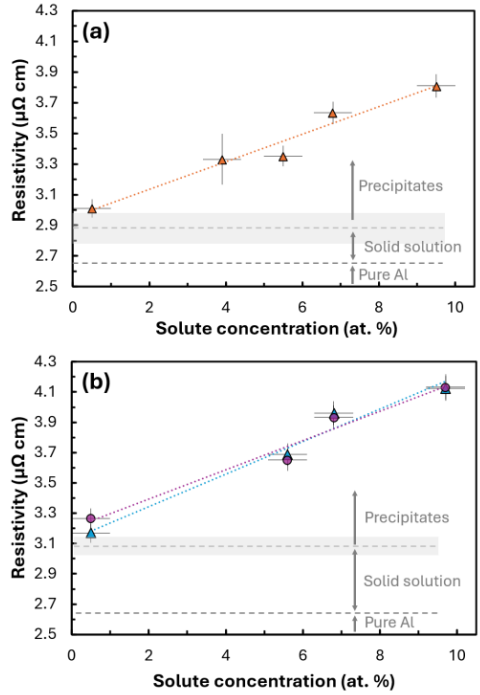

**Figure 7**: Bulk electrical resistivity values of Al-Mg-Si samples heat treated at (a) 350°C for 72 h and (b) 300°C for 24 h (blue) and 350°C for 120 h (purple). Solute concentration is the total concentration of Mg and Si. Uncertainties in concentration (error bars) are considered as 1 at. %.

It must be noted that equations (1) and (2) are equivalent, however the conventional specific resistivities of microstructural elements are substituted with an effective value. The effective specific resistivity refers to scattering by precipitates, while the physical quantity is solute content, which is more practical and straightforward value to determine.

**Table 2**: Coefficients of the linear regression analysis carried out for electrical resistivity values vs. solute concentration in the Al-Mg-Si alloys, heat treated at different conditions.

| Heat treatment | $a$ ($\mu\Omega$ cm (at. %)$^{-1}$) | $b$ ($\mu\Omega$ cm) | $R^2$ |
|---|---|---|---|
| 300°C, 24 h | 0.11 ± 0.01 | 3.10 ± 0.09 | 0.972 |
| 350°C, 72 h | 0.09 ± 0.01 | 2.96 ± 0.07 | 0.954 |
| 350°C, 120 h | 0.10 ± 0.01 | 3.20 ± 0.09 | 0.966 |

## 4 Conclusions

To conclude, we quantitatively analyze the contribution of Si and $Mg_2Si$ precipitates in Al-Mg-Si alloys to electrical resistivity relying on microscale four-point probe measurements; while other microstructural defects are neglected due to their sufficiently low density. We find that the resistivity of Al solid solution grains, including nano-precipitates, depends on the heat treatment rather than composition. Specifically, similar treatments for different compositions yield the same resistivity of the Al matrix grain interior. Also, we show that the secondary phases, Si and $Mg_2Si$, contribute equally to resistivity. We provide direct proof that electrical resistivity of Al-Mg-Si alloys linearly scales with the solute content (Si and Mg), where the linear coefficients depend on the heat treatment. Our study demonstrates a novel route to characterize the material's electrical resistivity down to micron length scales, thus enabling to resolve the bulk resistivity to microstructural components such as precipitates.

## CRediT authorship contribution statement

**Gautam Kumar**: Writing – review & editing, Formal analysis, Methodology. **Amram Azulay**: Writing – original draft, Writing – review & editing, Formal analysis. **Omer Coriat**: Methodology. **Hanna Bishara**: Writing – original draft, Writing – review & editing, Formal analysis, Methodology, Conceptualization, Supervision, Funding acquisition.

## Data availability

Data will be made available upon reasonable request.

## Declaration of Competing Interest

The authors declare that they have no known competing financial interests or personal relationships that could have appeared to influence the work reported in this paper.

**Acknowledgements**

H.B. acknowledges support for this research by NOGA Ltd. and the Israel Science Foundation (ISF) Grant No. 1309/23. Authors acknowledge Dr. Davide Levy from the Jan Koum Center for Nanoscience and Nanotechnology, Tel Aviv University, for the X-ray diffraction measurements.